\documentclass[aps,floatfix,twocolumn,showpacs]{revtex4}
\usepackage{natbib}
\usepackage{dcolumn}
\usepackage{graphicx}

\newcommand{\sr}{{\rm sr}}
\newcommand{\cals}{\zeta_s}
\newcommand{\calr}{\zeta_r}
\newcommand{\onediag}{{\bf 1}}
\newcommand{\Mpc}{{\rm Mpc}}

\newcommand{\WMAP}{{\slshape WMAP~}}
\newcommand{\WMAPc}{{\slshape WMAP}}

\newcommand{\apjl}{Astrophys.~J.~Lett.}
\newcommand{\mnras}{Mon.~Not.~R.~Astron.~Soc.}
\newcommand{\aap}{Astron.~Astrophys.}
\newcommand{\aj}{Astron.~J.}
\newcommand{\physrep}{Phys.~Rep.}
\newcommand{\araa}{Ann.~Rev.~Astron.~Astrophys.}
\newcommand{\apjs}{Astrophys.~J.~Supp.}

\begin{document}

\title{Weak Lensing and CMB: Parameter forecasts including a running spectral index} 

\author{Mustapha Ishak}
\email{mishak@princeton.edu}

\author{Christopher M. Hirata}

\author{Patrick McDonald}

\author{Uro\v s Seljak} 

\affiliation{Department of Physics, Jadwin Hall, Princeton University,
  Princeton, NJ 08544, USA} 

\date{\today}

\begin{abstract}
We use statistical inference theory to explore the constraints from 
future galaxy weak lensing (cosmic shear) surveys 
combined with the current CMB constraints on cosmological parameters,
focusing particularly on the running of the spectral index 
of the primordial scalar power spectrum, $\alpha_s$. 
Recent papers have drawn attention to the possibility of measuring $\alpha_s$
by combining the CMB with galaxy clustering and/or the Lyman-$\alpha$ forest.  
Weak lensing combined with the CMB provides an alternative
probe of the primordial power spectrum. 
We run a series of simulations with variable runnings and compare them 
to semi-analytic non-linear mappings to
test their validity for our calculations. We find that a ``Reference''
cosmic shear survey with $f_{sky}=0.01$ and $6.6\times 10^8$ galaxies
per steradian can reduce the uncertainty on $n_s$ and $\alpha_s$ by roughly a 
factor of 2 relative to the CMB alone. We investigate  the effect of
shear calibration biases on lensing by including the calibration
factor as a parameter, and show that for our Reference Survey, the
precision of cosmological parameter determination is only slightly
degraded even if the amplitude calibration is uncertain by as much as
5\%. We conclude that in the near future weak lensing surveys can
supplement the CMB observations to constrain the primordial power 
spectrum. 
\end{abstract}


\maketitle

\section{Introduction}
\label{sec:intro}

The recent observations of the cosmic microwave background (CMB)
from the {\slshape Wilkinson Microwave Anisotropy Probe} (\WMAPc)
mission confirmed the standard cosmological model to a very high
degree of accuracy \cite{2003ApJS..148..175S}.
The \WMAP data alone is consistent with a spatially flat model
dominated by dark energy and dark matter, with nearly scale-invariant,
adiabatic and Gaussian initial perturbations consistent with the
simplest inflationary models
\cite{1981PhRvD..23..347G,1982PhLB..108..389L,1982PhRvL..48.1220A}.

One of the remaining outstanding issues is whether the shape of the 
primordial power spectrum is consistent with the theoretical 
predictions.
This is usually expressed in terms of the slope of the primordial 
scalar spectrum $n_s$ and its running $\alpha_s=dn_s/d\ln k$. 
Most of the existing models predict that the slope is close, but 
not identical, to 
scale invariant, $n_s \sim 1$, and that running is small, 
$\alpha_s \sim (1-n_s)^2$. The best way to observationally settle 
this question is by combining data over a wide range of scales.
Combining data from \WMAPc, the Cosmic Background Imager (CBI)
\cite{2002AAS...200.0606P}, and the Arcminute Cosmology Bolometer Array 
Receiver (ACBAR)
\cite{2002AAS...20114004K}, together with constraints from the Two
Degree Field Galaxy Redshift Survey (2dFGRS) \cite{2001MNRAS.328.1039C}
Ref. \cite{2003ApJS..148..175S} found 0.05 error on $n_s$ 
and 0.025 on $\alpha_s$, with scale invariant model fitting the data at 
1.3 $\sigma$. 
Ref. \cite{2003ApJS..148..175S}
added Lyman-$\alpha$ forest constraints from
\cite{2002ApJ...581...20C,2002MNRAS.334..107G} and reduced the error 
on running to 0.016 (as well as found evidence for
running at $1.9\sigma$), but 
Ref. \cite{2003MNRAS.342L..79S} showed that the assumed errors were 
too small and 
the current Lyman-$\alpha$ forest constraints do not add 
significantly to the constraints from the CMB.   
The constraints on $\alpha_s$ in \cite{2003ApJS..148..175S} were obtained assuming no tensor modes,
 the strong degeneracy between running and tensors in the current data
 complicates the
interpretation of the one-dimensional marginalized probability
distribution for $\alpha_s$ \cite{2003MNRAS.342L..79S}.
Other authors \cite{2003MNRAS.342L..72B,2003astro.ph..6305L,2003hep.ph....2150B,2003hep.ph....5130K} who did not  
use Lyman-$\alpha$ forest, also find no significant evidence for a running 
spectral index, with errors at the level of 0.03.

While the current data show no evidence for running, the errors are still 
large and will be improved in the future. The most immediate improvement 
will come from the Sloan Digial Sky Survey (SDSS)\cite{2000AJ....120.1579Y}. The galaxy power 
spectrum analysis is not expected to improve the 2dF results 
in a statistical sense (but will be an important check of systematics), 
since on scales smaller than a few Mpc the nonlinear bias 
becomes intractable. On the other hand,
the Lyman-$\alpha$ forest spectrum and bispectrum analysis of several thousand 
quasar spectra in SDSS will 
lead to an order of magnitude improvement of the Lyman-$\alpha$ constraints
\cite{2003MNRAS.344..776M}. In combination with the CMB this is expected to 
give an error of $5\times 10^{-3}$ on $\alpha_s$. 

There are however 
uncertanties associated with the baryonic physics of the Lyman-$\alpha$
forest.
Future high-$\ell$ CMB data from the {\slshape Planck} satellite
(and, to a lesser extent, measurement of the second and third
acoustic peaks from future \WMAP data) will enable precision measurement of $\alpha_s$
independently of the galaxy and Lyman-$\alpha$ data,
but contamination from secondary anisotropies and foregrounds at high $\ell$
may be significant, since we are interested
in effects of only a few percent.  (The measurement of the first acoustic
peak by \WMAP is comparatively clean due to the large cosmological signal.)

It is therefore desirable to have yet another
probe of the primordial power spectrum at high wavenumbers.
Weak lensing is an attractive candidate since most of the potential
systematics are related to the difficulty of the observations and are
unrelated to the systematics associated with the other cosmological
probes we have discussed.
The purpose of this paper is to explore the constraints from future galaxy weak
lensing (cosmic shear) surveys combined with the current CMB results on
cosmological parameters, particularly the running of the scalar spectral 
index. 
Weak lensing is one of the most promising tools for an era of precision
cosmology as it probes directly the distribution of the
gravitating matter and does not suffer from the uncertainties associated with
nonlinear bias or gas physics
(for reviews, see
\cite{2003astro.ph..7212R,2001PhR...340..291B,1999ARA&A..37..127M} and
references therein). 

A few authors have used statistical inference theory to report forecasts on 
how well cosmic shear surveys combined with CMB data are able to constrain
cosmological parameters. See, for example
\cite{1999ApJ...514L..65H,2002PhRvD..65b3003H}. Our goal is to extend 
these studies to include the running, as well as investigate the 
effects of possible systematics on the results. More recently,  
Refs. \cite{2003PhRvL..90v1303C,2002astro.ph.12417W} combined CMB data
from various experiments with the Red-Sequence Cluster Survey (RCS)
weak lensing survey data (53 deg$^2$)
\cite{2002ApJ...577..595H,2002ApJ...572...55H}. Even with a modest 
survey, they found that their analysis 
reduces the uncertainties by comparable amounts or better than when
the CMB results are combined with other types of probes because it breaks
degeneracies present in the CMB data.  Their results 
are consistent with the earlier forecasts.
Other recent papers that reported statistical forecasts on how well 
weak lensing will be able to constrain various cosmological parameters
include 
\cite{2002PhRvD..65f3001H,2003PhRvL..91d1301A,2003astro.ph..6033B,2003astro.ph.10125T,
2003astro.ph.11104T,2003MNRAS.343.1327H,2003PhRvL..91n1302J,2003astro.ph..9332B,2003astro.ph..9032S}.

The potential of weak lensing as a cosmological probe has been realized quickly and there are many ongoing, planned,
and proposed surveys, such as the Deep Lens Survey ({\slshape http://dls.bell-labs.com/})
\cite{2002SPIE.4836...73W}; the NOAO Deep Survey ({\slshape http://www.noao.edu/noao/noaodeep/}); the
Canada-France-Hawaii Telescope (CFHT) Legacy Survey ({\slshape http://www.cfht.hawaii.edu/Science/CFHLS/})
\cite{2001misk.conf..540M}; the Panoramic Survey Telescope and Rapid Response System ({\slshape
http://pan-starrs.ifa.hawaii.edu/}); the {\slshape Supernova Acceleration Probe} ({\slshape SNAP}; {\slshape
http://snap.lbl.gov/}) \cite{2003astro.ph..4417R,2003astro.ph..4418M,2003astro.ph..4419R}; and the Large Synoptic
Survey Telescope (LSST; {\slshape http://www.lsst.org/lsst\_home.html}) \cite{2002SPIE.4836...10T}.  In this paper we
will discuss parameter constraints possible with a Reference Survey obtaining the shapes of $\bar{n} = 6.6\times
10^8$ galaxies per steradian over a fraction $f_{sky}=0.01$ of the sky (413 deg$^2$), with a peak redshift $z_p=1$.  
We also explore the effect of varying $f_{sky}$ and $\bar{n}$.  
The Reference Survey is somewhat more ambitious than the CFHT wide synoptic survey (172 deg$^2$); roughly
comparable to the {\slshape SNAP} wide survey; and significantly less ambitious than the LSST.

\section{Formalism and model}
\label{sec:form}

\subsection{Model parameters}

We consider the following basic parameter set for weak lensing:
$\Omega_{m}h^2$, the physical matter density;
$\Omega_\Lambda$, the fraction of the critical density in a
cosmological constant; $n_s(k_0)$ and $\alpha_s$, the spectral
index and running of the primordial scalar power spectrum at $k_0$;
$\sigma_8^{\rm lin}$, the amplitude of linear fluctuations; $z_p$, the
characteristic redshift of the source galaxies (see
Eq. \ref{eq:z_dist});  and $\cals$ and $\calr$, the calibration
parameters (see \S\ref{sec:calib}).  In order to combine this with information
from the CMB we include $\Omega_{b}h^2$, the physical baryon density;
$\tau$, the optical depth to reionization; $T/S$ the scalar-tensor
fluctuation ratio. We assume a spatially flat Universe with
  $\Omega_{m}+\Omega_{\Lambda}=1$, thereby fixing $\Omega_m$ and $H_0$
as functions of our basic parameters, and we do not include dynamical
dark energy, massive neutrinos, or primordial isocurvature
perturbations.  The primordial power spectrum of scalar density
fluctuation is given by \cite{1995PhRvD..52.1739K} 
\begin{equation}
P(k)=P(k_0)\left(\frac{k}{k_0}\right)^{n_{s}(k_{0})+\frac{1}{2}\alpha_{s}\ln\left(\frac{k}{k_{0}}\right)}, 
\label{eq:mps}
\end{equation}
where $n_{s}(k) \equiv \frac{d\ln P}{d \ln k}$ and $\alpha_{s}(k) \equiv
\frac{d\ln n_{s}}{d \ln k}$.  We assume $\frac{d^2\ln n_{s}}{d \ln
k^2}=0$, i.e. we ignore higher-order terms in the Taylor expansion of
the primordial power spectrum.  We have taken our pivot wavenumber to
be $k_0=0.05$ h/Mpc. We use as fiducial model 
the best fit for the WMAP(1yr)+ACBAR+CBI data from the Markov Chains in 
Ref. \cite{2003MNRAS.342L..79S}: $\Omega_b
h^2=0.0228$, $\Omega_m h^2=0.139$, $\Omega_{\Lambda}=0.74$,
$n_s=0.95$, $\alpha_s=0$ (with $\pm0.04$ variations),
$\sigma_{8}=0.85$, $\tau=0.177$, $T/S=0.265$, $z_p=1.0$, $\cals=0.0$,
and $\calr=0.0$.

\subsection{Convergence power spectrum}

We use the Limber approximation to the convergence power spectrum
\cite{1992ApJ...388..272K,1997ApJ...484..560J,1998ApJ...498...26K},
which is valid on small angular scales $\ell\gg 1$ for which the
gravitational lensing deflection can be approximated as a random walk
due to many independent structures along the line of sight: 
\begin{equation}
P_{\kappa}(\ell) =
\frac{9}{4} H_0^4\Omega_m^2\int^{\chi_H}_{0}
\frac{g^2(\chi)}{a^2(\chi)}P_{3D}\left(\frac{\ell}{\sin_{K}(\chi)},\chi\right) d\chi.
\end{equation}
Here $P_{3D}$ is the $3D$ nonlinear power spectrum of the matter
density fluctuation, $\delta\rho/\rho$; $\chi$ is 
the radial comoving coordinate; $a(\chi)$ is the scale factor; and 
$\sin_{K}\chi=K^{-1/2}\sin(K^{1/2}\chi)$ is 
the comoving angular diameter distance to $\chi$.
The weighting function $g(\chi)$ is the source-averaged distance ratio 
given by 
\begin{equation}  
g(\chi) = \int_\chi^{\chi_H} n(\chi') {\sin_K(\chi'-\chi)\over
\sin_K(\chi')} d\chi',
\end{equation}
where $n(\chi(z))$ is the source redshift distribution normalized by
$\int dz\; n(z)=1$.

\begin{figure}[ht]
\includegraphics[width=2.4in,angle=-90]{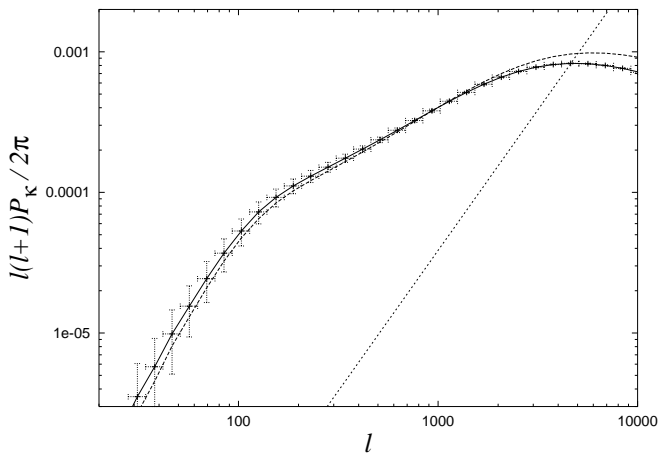}
\caption{\label{fig:fig1} Curved convergence power
  spectra for the Reference Survey with $\alpha_s=-0.04$ (solid curve) and
  $\alpha_s=0.04$ (dashed); $k_0=0.05 h/Mpc$ and all other parameters
  are fixed at their fiducial   values. The dotted line represent the
  noise power spectrum.   The Reference Survey is sampling
  variance-limited out to $\ell\approx 2000$.} 
\end{figure}

For our weak lensing calculations, we used the BBKS linear transfer function \cite{1986ApJ...304...15B}
appropriate for cold dark matter with adiabatic initial perturbations.
No baryonic correction was applied, since $\Omega_bh^2$ is tightly constrained
by the CMB and since the baryonic oscillations that appear in the full
transfer function are smoothed out by projection effects when the convergence
power spectrum is determined.  We have used the analytic approximation to
the growth factor of Ref. \cite{1991MNRAS.251..128L}.
We used the recent nonlinear mapping procedure {\sc halofit}
\cite{2003MNRAS.341.1311S} to compute 
the non-linear power spectrum. This procedure, as described in
\cite{2003MNRAS.341.1311S}, is based on a fusion of the halo model and
an HKLM scaling \cite{1991ApJ...374L...1H} and is more accurate than
the commonly used Peacock-Dodds mapping \cite{1996MNRAS.280L..19P}.
We discuss it further in section \ref{sec:simulations} and check it
versus numerical simulations.
We contrast in Fig. \ref{fig:fig1} two curved convergence power spectra
for $\alpha_s=-0.04,0.04$ and display the sample variance errors
averaged over bands in $\ell$ and the noise in the measurement.  
For $n(z)$, we take the fitting function
\cite{2000Natur.405..143W}:
\begin{equation}
n(z)=\frac{z^2}{2 z_0^3}\, e^{-z/z_0},
\label{eq:z_dist}
\end{equation}
which peaks at $z_p=2z_0$.
When we consider tomography,
the sources are split in two
bins: $z<z_p$ and $z\ge z_p$.  The normalized source redshift distributions for these two bins are:
\begin{eqnarray}
{n_{A}(z)= \frac{z^{2}}{2 z_{0}^{3}(1-5/e^{2})}\,{e^{-z/z_{0}}}} &
  {\;\;{\rm for}\;\;} {z_{p}< 2 z_{0}} \nonumber \\ 
{n_{B}(z)= \frac{e^2 z^2}{10 z_0^3 }\, {e^{-z/z_0}} } & \;\;{\rm for}\;\;
  {z_p\ge 2 z_0}.
\label{eq:z_dist2}
\end{eqnarray} 
\subsection{Calibration parameters}
\label{sec:calib}

Weak lensing surveys today are subject to various systematic errors
that are comparable to the statistical errors.  One 
example is the shear calibration bias
\cite{2001A&A...366..717E, 2001MNRAS.325.1065B, 2003MNRAS.343..459H,
2002AJ....123..583B, 2003astro.ph..5089V, 2000ApJ...537..555K}, in which
the gravitational shear is systematically over- or under-estimated by
a multiplicative factor.  Physically, the principal source of 
this bias is incomplete correction for the circularization of images
of galaxies by the point-spread function (PSF), although there are 
also noise- and selection-related contributions.
The shear calibration bias is
particularly dangerous because the usual systematic error tests
applied in weak lensing -- e.g. decomposition of the shear field into 
$E$ and $B$ modes, cross-correlation of the shear map against maps of
the PSF, etc. -- are completely 
insensitive to this bias.  Indeed, shear calibration bias mimics an
overall rescaling of the shear power spectrum, and proposals to 
circumvent it have thus far been based on detailed simulations of the
observation. 

We have parameterized the shear calibration bias here using the power
calibration parameter $\cals$; that is, the measured convergence 
power spectrum $\hat P_\kappa(\ell)$ is given by:
\begin{equation}
\hat P_\kappa(\ell) = (1+\cals) P_\kappa(\ell),
\label{eq:cals}
\end{equation}
where $P_\kappa(\ell)$ is the power spectrum obtained in the absence
of calibration errors.  Note that $\cals$ refers to the calibration 
error of the power spectrum, which is twice the calibration error of
the amplitude because the power spectrum is proportional to 
amplitude squared.

When we consider tomography, we must also consider
the relative calibration $\calr$ between the two redshift
bins.  This error affects the measured power spectrum $\tilde
P_\kappa(\ell)$ in accordance with:
\begin{eqnarray}
\tilde P_\kappa^{AA} = && \!\!\!\! (1+f_B \calr) \hat P_\kappa^{AA}(\ell), 
\nonumber \\
\tilde P_\kappa^{AB} = && \!\!\!\! (1+{f_B-f_A\over 2} \calr) \hat P_\kappa^{AB}(\ell), 
\nonumber \\
\tilde P_\kappa^{BB} = && \!\!\!\! (1-f_A \calr) \hat P_\kappa^{BB}(\ell), 
\label{eq:calr}
\end{eqnarray}
where $f_A\approx 0.32$ is the fraction of the source
galaxies in bin A and $f_B\approx 0.68$ is the fraction
of the source galaxies in bin B.
Mathematically, this means that if
$\calr=0.01$, then the power spectrum calibrations of the two redshift 
bins are offset 1\% relative to each other, but the calibration is
correct when averaged over all the source galaxies.
Physically, $\calr$ parameterizes a redshift-dependent calibration bias,
which could arise e.g. from an incomplete PSF correction whose residual
error depended on the galaxy's angular size or magnitude (which correlates
with redshift).

It is possible for a redshift-dependent calibration bias to occur even without
tomography, and hence we have included the relative calibration bias $\calr$
even in our no-tomography parameter forecasts.  In this case, the measured
convergence power spectrum is computed from:
\begin{equation}
\hat P_\kappa(\ell) = f_A^2 \hat P_\kappa^{AA}(\ell) + 2f_Af_B 
\hat P_\kappa^{AB}(\ell)
+ f_B^2 \hat P_\kappa^{BB}(\ell),
\end{equation}
where the $\hat P_\kappa^{IJ}(\ell)$ are given by Eq. (\ref{eq:calr}).  In
principle the relative calibration bias can influence parameter estimation
because it alters the effective source
redshift distribution.  However, we have found that it only slightly affects
our parameter forecasts in the no-tomography case.

In a real experiment, $\cals$ and $\calr$ are not completely unknown,
but rather are parameters of the experiment that can in principle 
be determined by simulating observations.  We have therefore imposed
Gaussian priors of width 
$\sigma_{pr}(\cals)$ and $\sigma_{pr}(\calr)$ on the calibration
parameters (for simplicity we have taken $\cals$ and $\calr$ to 
be uncorrelated in the case of tomography); here the $\sigma_{pr}$ are
the uncertainties in the shear 
calibration of the experiment.  The prior curvature matrix (with
diagonal elements $\sigma_{pr}^{-2}$ corresponding to the shear 
calibration parameters) is used in computing parameter
uncertainties in accordance with Eq. (\ref{eq:sigma}).

It is important to note that $\{\cals,\calr\}$ is far from a complete
parameterization of systematic errors.  Other effects 
include spurious power from non-circular PSF, intrinsic alignments of
galaxies, etc.  We have not investigated these here, although 
clearly they must be minimized and the residual errors estimated if
weak lensing is to evolve into a precision cosmological tool.

\section{Numerical simulations}
\label{sec:simulations}

The {\sc halofit} code described in \cite{2003MNRAS.341.1311S} is  
a physically motivated fitting formula to the results of 
numerical simulations.
It is not obvious whether it is accurate enough for the present purpose:
computing the derivatives of the non-linear power spectrum
that are used to form the Fisher matrix.  Ref. \cite{2003MNRAS.341.1311S}
effectively varied all the parameters we use here, but their sampling of
parameter space may not have been sufficiently dense to constrain the 
interpolation near our point of interest.
We performed a set of particle-mesh $N$-body simulations to check
{\sc halofit} in this context, using the Tree-Particle-Mesh (TPM) code described in
\cite{2003ApJS..145....1B}. 
Each of our parameters is explicitly varied around the
central model and the changes in the power spectra are compared to the
predictions of the fitting formula (the figures we show
are for the power at $z=0.5$).  The central model and variations are 
$\sigma_8=0.85\pm 0.05$, $\Omega_\Lambda=0.74\pm 0.04$, 
$\Omega_m h^2=0.139\pm 0.01$, $n=0.95\pm 0.05$, and $\alpha=0.0\pm0.04$.
We always compare $\Delta P/P$, the fractional change in power, because
most of the statistical error in $\Delta P$ is removed this way and the
convergence with resolution is better.  When we check how the simulations
affect the Fisher matrix, we compute it using the formula 
$\Delta P = (\Delta P_{sim} / P_{sim}) P_{fit}$, where $P_{sim}$ means
power from the simulation, and $P_{fit}$ means power from {\sc halofit}.

\begin{figure}[ht]
\includegraphics[width=2.4in]{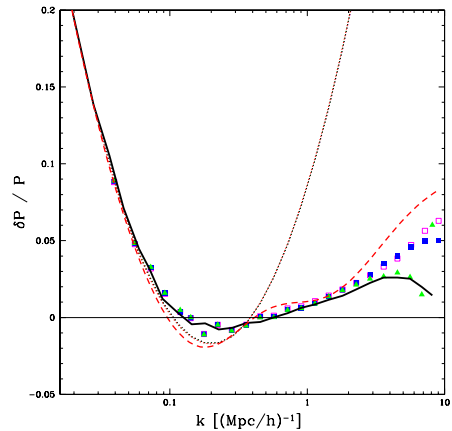}
\caption{\label{fig:simalpha}
The fractional difference between the nonlinear mass power spectra 
at $z=0.5$ for
$\alpha(k_0=0.2 h/\Mpc)=0.04$ and $\alpha(k_0=0.2 h/\Mpc)=-0.04$,
$[P(k,\alpha=+0.04)-P(k,\alpha=-0.04)]/\bar{P}(k)$, where 
$\bar{P}(k)=[P(k,\alpha=+0.04)+P(k,\alpha=-0.04)]/2$.
All other parameters are fixed at their fiducial values. 
The dotted line is linear theory
The solid line is from an $L=320 h^{-1}\Mpc$, $N=256^3$ simulation.
The open squares, filled squares, and triangles are from $L=160 h^{-1}\Mpc$ 
simulations with
$N=512^3$, $N=256^3$, and $N=128^3$, respectively.  Note that the triangle at the 
highest k is well beyond the resolution limit of the simulations.
The dashed line is from {\sc halofit}. 
}
\end{figure}

Figure \ref{fig:simalpha} shows the effect of 
changing $\alpha$, for a pivot point $k_0=0.2 h/\Mpc$ that roughly corresponds
to the scale of $\sigma_8$ and weak lensing.
To be sure that our results have numerically converged, we compare simulations
with box size $L=320 h^{-1}\Mpc$ and $N=256^3$ particles (our mesh for the force
calculation is always a factor of 2 finer than the mean particle spacing)
to simulations with $L=160 h^{-1}\Mpc$ and $N=128^3$ (the same resolution as the
bigger box), $N=256^3$, or $N=512^3$.  
In Fig. \ref{fig:simalpha} we note first that the
difference between the two box sizes is probably insignificant (solid line vs. 
triangles).  The $L=160 h^{-1}\Mpc$, $N=256^3$ box has
sufficient resolution to compute derivatives accurately out to 
$k=10 h/\Mpc$ (open vs. closed squares).
Turning to the comparison between {\sc halofit} and the simulations, we see
that the fitting formula does well if one is not interested in the
fine details of the nonlinear power; however, a Fisher matrix constraint that 
actually relies on the value of the derivative at $k\sim 0.2 h/\Mpc$ (rather
than just the fact that it is close to zero) would be inaccurate.
\begin{figure}[ht]
\includegraphics[width=2.4in,angle=-90]{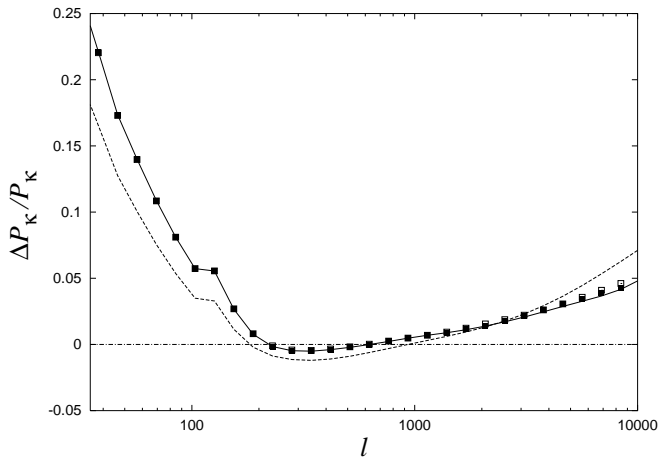}
\caption{\label{fig:comparepkappa}
The fractional difference between the convergence power spectra 
for
$\alpha(k_0=0.2 h/\Mpc)=0.04$ and $\alpha(k_0=0.2 h/\Mpc)=-0.04$,
$[P_{\kappa}(l,\alpha=+0.04)-P_{\kappa}(l,\alpha=-0.04)]/\bar{P_{\kappa}}(l)$, where 
$\bar{P_{\kappa}}(l)=[P_{\kappa}(l,\alpha=+0.04)+P_{\kappa}(l,\alpha=-0.04)]/2$.
All other parameters are fixed at their fiducial values. 
The solid line is from an $L=320 h^{-1}\Mpc$, $N=256^3$ simulation.
The open squares and filled squares are from $L=160 h^{-1}\Mpc$ 
simulations with
$N=512^3$  and $N=256^3$ respectively.  
The dashed line is from {\sc halofit}. 
}
\end{figure}
We show in \ref{fig:comparepkappa} 
how the differences in the {sc halofit} and n-body mass power spectra are
transferred to the convergence power spectra.

\begin{figure}[ht]
\includegraphics[width=2.4in]{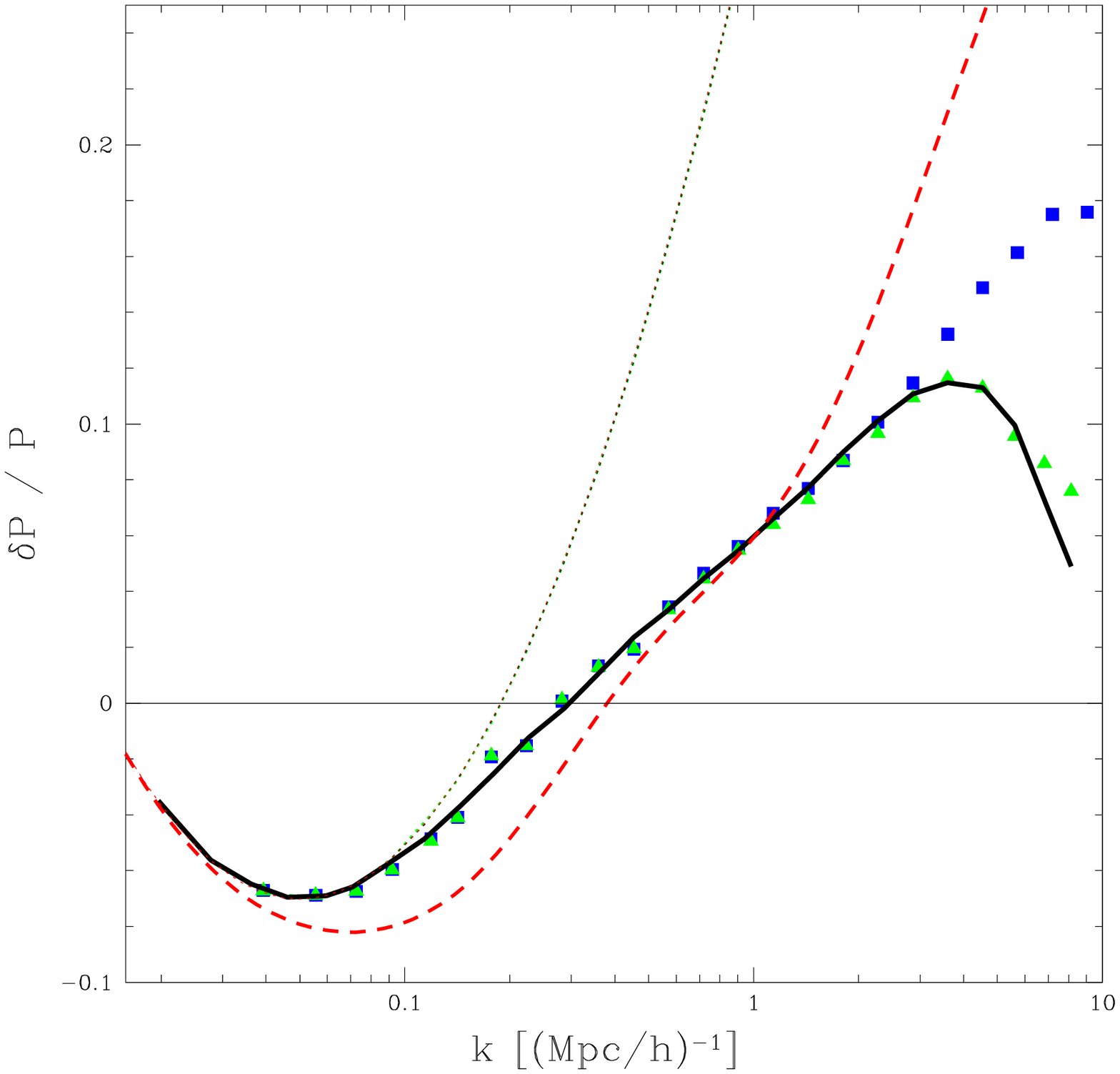}
\caption{\label{fig:simalphak05} 
The fractional difference between the nonlinear mass power spectra 
at $z=0.5$ for
$\alpha(k_0=0.05 h/\Mpc)=0.04$ and $\alpha(k_0=0.05 h/\Mpc)=-0.04$.
The dotted line is linear theory.
The solid line is from an $L=320 \Mpc/h$, $N=256^3$ simulation.
The squares and triangles are from $L=160 \Mpc/h$ 
simulations with $N=256^3$ and $N=128^3$, respectively.  
The dashed line is from {\sc halofit}.
}
\end{figure}

\begin{figure}[ht]
\includegraphics[width=2.4in]{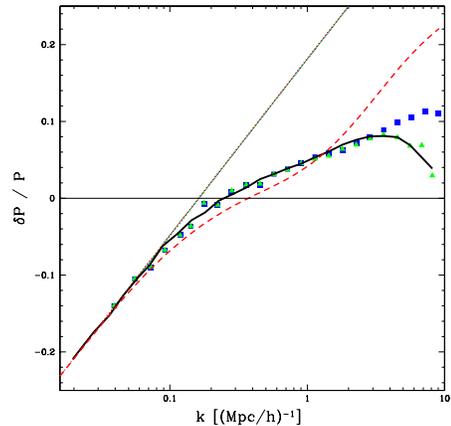}
\caption{\label{fig:simenn} 
The fractional difference between the nonlinear mass power spectra 
at $z=0.5$ for
$n=1.0$ and $n=0.9$.
The dotted line is linear theory.
The solid line is from an $L=320 \Mpc/h$, $N=256^3$ simulation.
The squares and triangles are from $L=160 \Mpc/h$ 
simulations with $N=256^3$ and $N=128^3$, respectively.  
The dashed line is from {\sc halofit}.
}
\end{figure}

For the combination of CMB and lensing we use the pivot point 
$k_0=0.05 h/\Mpc$.  Figure \ref{fig:simalphak05} shows the comparison
between the simulations and {\sc halofit} for this $k_0$. 
The result is similar to the $k_0=0.2 h/\Mpc$ case, i.e., the basic 
trend is correct but the details are not completely correct.
The same could be said of the results for variation in $n$, shown in
Fig. \ref{fig:simenn}.

\begin{figure}[ht]
\includegraphics[width=2.4in]{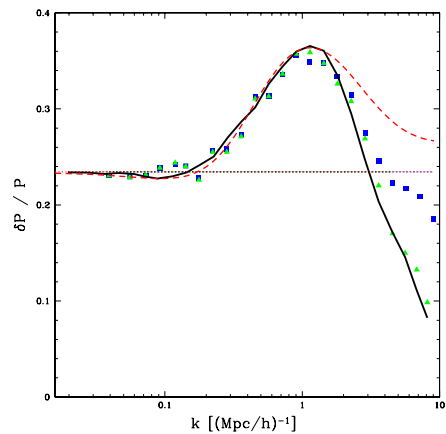}
\caption{\label{fig:simsigma8} 
The fractional difference between the nonlinear mass power spectra 
at $z=0.5$ for
$\sigma_8=0.9$ and $\sigma_8=0.8$.
See Fig. \ref{fig:simenn} for line meanings.
}
\end{figure}

\begin{figure}[ht]
\includegraphics[width=2.4in]{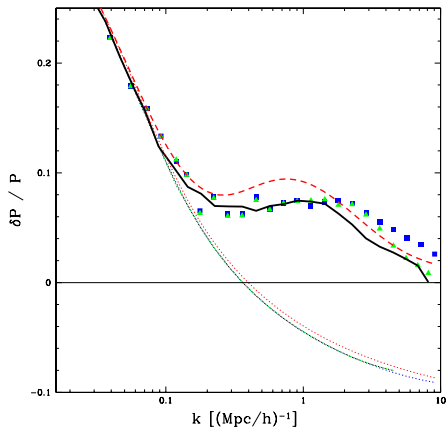}
\caption{\label{fig:simOL} 
The fractional difference between the nonlinear mass power spectra 
at $z=0.5$ for
$\Omega_\Lambda=0.78$ and $\Omega_\Lambda=0.7$.
See Fig. \ref{fig:simenn} for line meanings.
}
\end{figure}

\begin{figure}[ht]
\includegraphics[width=2.4in]{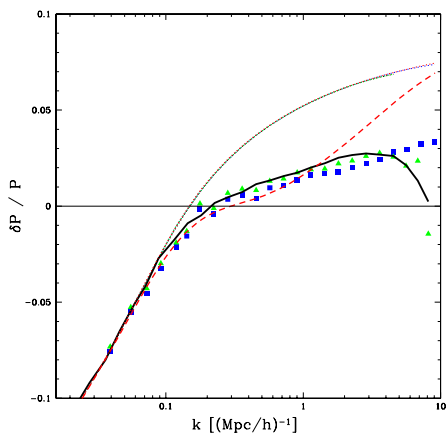}
\caption{\label{fig:simoM} 
The fractional difference between the nonlinear mass power spectra 
at $z=0.5$ for
$\Omega_m h^2=0.149$ and $\Omega_m h^2=0.129$.
See Fig. \ref{fig:simenn} for line meanings.
}
\end{figure}

The comparisons for variations in $\sigma_8$, $\Omega_\Lambda$,
and $\Omega_m h^2$ are shown in Figures \ref{fig:simsigma8},
\ref{fig:simOL}, and \ref{fig:simoM}, respectively.  Generally the
agreement for these cases seems better (i.e., very good).  
For $k \lesssim 2 h/\Mpc$ the 
error in the prediction of the derivatives by the 
fitting formula is not large, even as a fraction of the values of
the derivatives.  

Note that our comparison of {\sc halofit} to the simulations has been restricted to the fractional
change between models, and implies nothing about the accuracy of {\sc halofit} for predicting the
absolute power spectrum, which is necessary when interpreting data.  Our simulations are not
suited to predicting the absolute power -- on large scales they have big statistical
fluctuations due to limited box size, while on small scales the power is suppressed by the
limited PM resolution.  Both of these effects cancel neatly in the fractional derivatives
that we show, and for our Fisher matrix calculation this is sufficient.


\section{Parameter forecasts}

\subsection{Fisher-matrix analysis}

The uncertainty in the observed weak lensing spectrum is
given by: \cite{1992ApJ...388..272K,1998ApJ...498...26K}
\begin{equation}
\Delta P_{\kappa}(\ell)=
\sqrt{\frac{2}{(2\ell +1)f_{sky}}}\left (
P_{\kappa}(\ell) + {\left< \gamma_{int}{}^2 \right>\over \bar n} \right ) \,,
\label{delta_kappa}
\end{equation}
\noindent where $f_{sky} = \Theta^2 \pi/129600 $ is the fraction of the
sky covered by a survey of dimension $\Theta$ in degrees, 
$\left<\gamma_{int}^2\right>^{1/2} \approx 0.4$ is the intrinsic
ellipticity of galaxies. We assume for a Reference Survey a sky coverage
of $f_{sky}=0.01$ and an average galaxy number density of
$\bar n \approx 6.6 \times 10^{8} \sr^{-1}$; we will also investigate
the effects of varying both of these parameters.

The Fisher-matrix formalism for cosmological parameter forecast has
been proven in previous studies to be a powerful tool for estimating
the statistical errors achievable by experiments 
\cite{1999ApJ...518....2E,1997ApJ...488....1Z,1997MNRAS.291L..33B,1999ApJ...514L..65H,2002PhRvD..65f3001H}.
If the convergence field is Gaussian, and the noise is a combination
of Gaussian shape and instrument noise with no intrinsic 
correlations, the Fisher matrix is given by:
\begin{equation}
F_{\alpha \beta} = {\sum_{\ell=\ell_{\rm min}}^{\ell_{\rm
	max}}{\frac{(\ell + 1/2) f_{sky}}{(P_\kappa + \left< \gamma_{\rm
	int}^2\right > /\bar n)^2} {\frac{\partial P_\kappa}{\partial p^{\alpha}}}} 
	{\frac{\partial P_\kappa}{\partial p^{\beta}}}};
\label{eq:fisher1}
\end{equation}
we have used $\ell_{\rm max}=3000$ since on smaller scales, the assumption
of a Gaussian shear field underlying Eq. (\ref{eq:fisher1}) and the 
{\sc halofit} approximation to the nonlinear power spectrum may not be valid.
(Note that the shear field is a projection through many nearly independent
structures, so by the Central Limit Theorem it can be
well-described by Gaussian statistics even when the density perturbations are not
Gaussian.  Even with the Central Limit Theorem, deviations of the shear field
from Gaussian statistics become large at the 1--2 arcmin scale
\cite{2000ApJ...530..547J,2001ApJ...554...56C}.)
On small scales, there is cosmological information in the small-scale
non-Gaussianity (e.g. skewness) of the lensing field \cite{1997A&A...322....1B},
but we do not investigate this here.  For the minimum $\ell$, we take the fundamental
mode approximation:
\begin{equation}
\ell_{\rm min} \approx \frac{360\rm ~deg}{\Theta} = \sqrt{\pi\over f_{sky} },
\label{eq:lmin}
\end{equation}
i.e. we consider only lensing modes for which at least one wavelength can fit inside
the survey area.  The survey contains some information on larger angular scales, and for
this reason the approximation Eq. (\ref{eq:lmin}) might be considered conservative.
However, it should also be noted that some of the planned surveys will scan disconnected
regions of the sky in order to provide additional systematic error checks, in which
case the effective $\ell_{\rm min}$ is increased.

\begin{figure}[ht]
\includegraphics[width=2.4in,angle=-90]{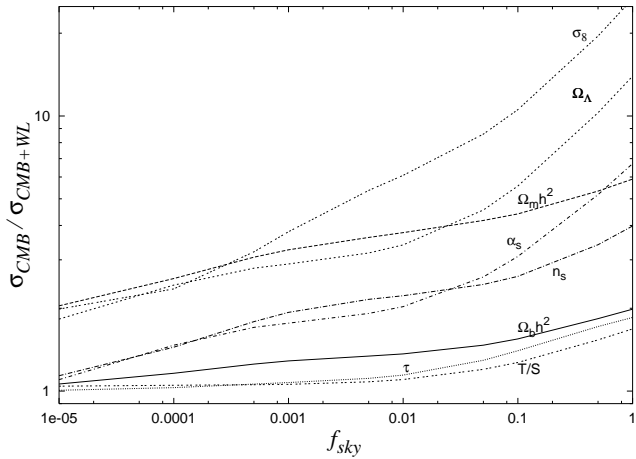}
\caption{\label{fig:fig3} Improvements of the CMB parameter estimation
with weak lensing. The effect of increasing $f_{sky}$ is displayed
here.}
\end{figure}

The statistical error on a given parameter $p_{\alpha}$ is then given
by:
\begin{equation}
\sigma^2(p_{\alpha})\approx [({\bf F}+\Pi)^{-1}]_{\alpha \alpha},
\label{eq:sigma}
\end{equation}
where $\Pi$ is the prior curvature matrix.  We only impose priors on
the source redshift and on the calibration parameters (see 
\S\ref{sec:calib}).  For the Reference Survey, we take priors of
$\sigma(\cals)=\sigma(\calr)=0.02$ on the calibration parameters
\cite{2003MNRAS.343..459H}
and $\sigma(z_p)=0.05$ on the source redshifts.
Since the primary CMB anisotropies are generated at much larger comoving
distance than the density fluctuations that give rise to weak lensing,
it is a good approximation to take them to be independent;
in this case, we can add the Fisher matrices from lensing and CMB to
yield combined constraints on cosmological parameters.
This combination leads to significant improvements in parameter estimation as we show
in Fig. \ref{fig:fig3}.

\subsection{Adding tomography}

\begin{figure}[ht]
\includegraphics[width=2.4in,angle=-90]{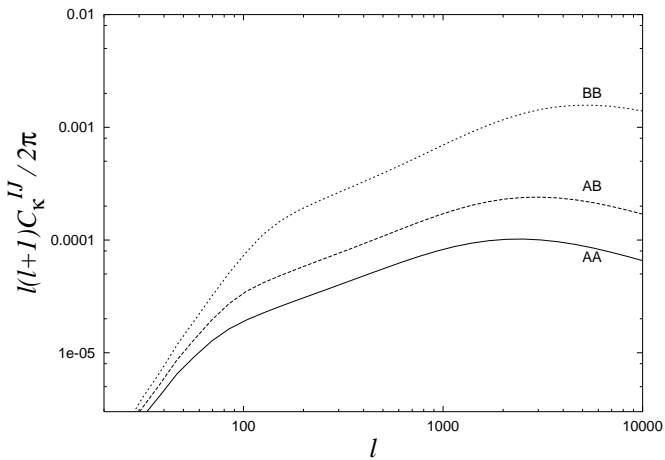}
\caption{\label{fig:fig5} Convergence power spectra for bin A, bin B
  and their cross correlation. $\alpha_s=-0.04$ and $k_0=0.05 h/Mpc$.}
\end{figure}

Tomography has been shown to improve significantly the measurements of
cosmological parameters \cite{1999ApJ...522L..21H,2002PhRvD..65b3003H}. We explore
the impact on parameter estimation from tomographic separation of the
source galaxies into two redshift bins; see Eq.~(\ref{eq:z_dist2}).  
The Fisher formalism is generalized here using:
\begin{equation}
 F_{\alpha\beta} = \sum_{\ell_{\rm min}}^{\ell_{\rm max}}
		(\ell + 1/2) f_{\rm sky}
 {\rm Tr}\left( {\bf C}_\ell^{-1} {\partial{\bf P}_\ell\over\partial p^\alpha} 
                       {\bf C}_\ell^{-1} {\partial{\bf P}_\ell\over\partial p^\beta} \right),
\label{eq:fisher2}
\end{equation}
where ${\bf C}_\ell$ is the covariance matrix of the multipole moments
of the observables ${C}^{\kappa \kappa'}_\ell = P_\ell^{\kappa \kappa'}
+ N_\ell^{\kappa \kappa'}$ with $N_\ell^{\kappa \kappa'}$ the power spectrum
of the noise in the measurement. These read
\begin{equation}
{P}^{\kappa \kappa'}_{\ell} = \left( \begin{array}{cc} 
P_\ell^{AA} & P_\ell^{AB} \\
P_\ell^{AB} & P_\ell^{BB} \end{array} \right),
\end{equation}
and
\begin{equation}
{N}^{\kappa \kappa'}_{\ell} = \left( \begin{array}{cc} 
{\left< \gamma_{\rm int}^2\right > /\bar n_{A}}  & 0 \\
0  & {\left< \gamma_{\rm int}^2\right > /\bar n_{B}} 
\end{array} \right).
\end{equation}
The convergence power spectra for the two bins and their cross
correlation are shown in Fig. \ref{fig:fig5}.

\subsection{Results}
\begin{table*}
\caption{\label{tab:contrast}A comparative table for parameter estimation
  errors from our Reference Survey with $f_{sky}=0.01$ and
  $\bar{n}=6.6 \times 10^8 \sr^{-1}$. The first row (C)
  are uncertainties from the CMB=\WMAPc (1yr)+CBI+ACBAR alone.
  The following rows show combined errors for CMB plus
  weak lensing in the Reference Survey (CW),
  and CMB plus weak lensing with tomography (CT).
  The lower-case suffixes a,b,c indicate the priors used:
  (a) prior of $0.1$ on $\cals$ and $\calr$ (if applicable),
  and $0.05$ on $z_p$; (b) prior of $0.02$ on $\cals$ and $\calr$)
  and $0.05$ on $z_p$; and (c) prior of $10^{-4}$ on $\cals$, $\calr$,
  and $z_p$ (i.e. effectively perfect knowledge of calibration and
  source redshift is assumed).
  For the rows labeled with [S], the $N$-body simulation was used
  to provide the matter power spectra instead of using {\sc halofit}.
  Note that the simulated and {\sc halofit} results are very similar.
}
\begin{ruledtabular}
\begin{tabular}{ccccccccccccc}
 &
$\sigma(\Omega_m h^2)$    &
$\sigma(\Omega_b h^2)$    & 
$\sigma(\Omega_\Lambda)$  &
$\sigma(\sigma_8)$        & 
$\sigma(n_s)$	          &
$\sigma(\alpha_s)$        &
$\sigma(\tau)$            &
$\sigma(T/S)$             &
$\sigma(z_p)$             &
$\sigma(\cals)$           &
$\sigma(\calr)$            \\
\hline
C      &0.013 &0.0012&0.054 & 0.083& 0.036& 0.039& 0.035 &0.154& -      & -    & -    \\
CWa    &0.0036&0.0010&0.0191& 0.024& 0.021& 0.017& 0.033 &0.143&   0.042 &0.092& 0.099 \\
CWb    &0.0030&0.0010&0.0167& 0.023& 0.021& 0.017& 0.033 &0.143&   0.041 &0.020& 0.020\\
CWc    &0.0027&0.0010&0.0145& 0.022& 0.021& 0.017& 0.033 &0.142&   0.0001&0.0001&0.0001\\
CTa    &0.0034&0.0010&0.0170& 0.021& 0.020& 0.016& 0.031 &0.140&   0.029 &0.075 &0.024\\
CTb    &0.0029&0.0009&0.0159& 0.017& 0.019& 0.015& 0.030 &0.138&   0.027 &0.020 &0.015\\
CTc    &0.0027&0.0008&0.0069& 0.010& 0.016& 0.013& 0.023 &0.130&   0.0001&0.0001&0.0001\\
\hline
CWb[S] &0.0032&0.0010&0.0169& 0.024& 0.022& 0.019& 0.034 &0.142&   0.045 &0.020 &0.020\\
CWc[S] &0.0027&0.0010&0.0151& 0.023& 0.022& 0.019& 0.034 &0.142&   0.0001&0.0001&0.0001\\
CTb[S] &0.0030&0.0009&0.0147& 0.017& 0.015& 0.014& 0.030 &0.125&   0.024 &0.020 &0.014\\
CTc[S] &0.0025&0.0008&0.0069& 0.010& 0.014& 0.013& 0.022 &0.123&   0.0001&0.0001&0.0001\\
\end{tabular}
\end{ruledtabular}
\end{table*}
We generated the convergence power spectra using a weak lensing
code that includes the formalism and features described in
\S\ref{sec:form}. We derived parameter uncertainties for weak lensing
with and without tomography and then combine these errors with the
CMB analysis. We used the CMB parameter estimation uncertainties of Ref.
\cite{2003MNRAS.342L..79S} (the correlation matrix was provided by the authors).
As described in their paper, the authors have used
a standard Monte Carlo Markov Chain using the
\WMAPc +CBI+ACBAR data (we will refer
to these as ``CMB''). In Table \ref{tab:contrast}, we show the
values for the uncertainties from CMB alone and then from a
combination of CMB with weak lensing with and without tomography for
our reference survey ($f_{sky}=0.01$ and  $\bar{n}=6.6 \times 10^8 \sr^{-1}$.)

We summarize in Table \ref{tab:varfsky} and Table \ref{tab:varn} the results respectively for increasing values of
$f_{sky}$ and $\bar{n}$.  We show the full correlation matrices for combined CMB and weak lensing observations with
and without tomography in Table \ref{tab:corrtab}. In Table \ref{tab:varfsky2}, we have fixed the calibration
parameters and the characteristic redshift of the source galaxies.  This significantly improves the precision of
cosmological parameter estimates; we can see from the correlation matrix (Table \ref{tab:corrtab}) that most of this
improvement comes from the fixing the source redshift.  As can be seen from Table \ref{tab:varfsky}, very stringent
cosmological constraints can in principle be obtained for surveys covering thousands of square degrees; however,
equally stringent controls over systematic errors would be required.  An additional caveat is the possibility of
errors in the parameter constraints at very high $f_{sky}$ (i.e. the bottom row of Tables \ref{tab:varfsky} and
\ref{tab:varfsky2}), since there the Fisher matrix is poorly conditioned and inaccuracies in our computation of the
derivatives $\partial P_\kappa(\ell)/\partial p^\alpha$ are thus magnified.

It is worth noting from the correlation matrix (Table \ref{tab:corrtab}) that the calibration parameters are not
degenerate with any of the cosmological parameters considered. This lack of degeneracy is good news because it means
that weak lensing surveys can be used for precision cosmology even if the absolute shear calibration cannot be
determined with high accuracy.  Note that the current data analysis methods for lensing are estimated to have
amplitude calibration correct at roughly the $\sim 5$\% level; the power calibration is twice this, or $\pm 0.1$
\cite{2003MNRAS.343..459H}. If we impose only this $\pm 0.1$ prior on $\cals$ and $\calr$ (and retain the $\pm 0.05$
prior on $z_p$), we see that the uncertainties on cosmological parameters are essentially unchanged (Table
\ref{tab:contrast}).  With this weaker prior, the correlation of the calibration parameters with $z_p$, 
$\sigma_8$, and $\Omega_\Lambda$ becomes larger ($|\rho|>0.5$), and with even weaker priors the uncertainties in 
these parameters are degraded.

\begin{table*}
\caption{\label{tab:varfsky}Parameter estimation errors for weak lensing
  combined with CMB: $f_{sky}$ is varied from
  $10^{-5}$ to 1.0. and  $\bar{n}=6.6 \times 10^8 \sr^{-1}$. We use priors of 0.02 on
  $\cals$ and $\calr$ and 0.05 on $z_p$. Tomography is added in the
  second part of the table.}
\begin{ruledtabular}
\begin{tabular}{cccccccccccc}
$f_{sky}$&
$\sigma(\Omega_m h^2)$    &
$\sigma(\Omega_b h^2)$    & 
$\sigma(\Omega_\Lambda)$  &
$\sigma(\sigma_8)$        & 
$\sigma(n_s)$	          &
$\sigma(\alpha_s)$        &
$\sigma(\tau)$            &
$\sigma(T/S)$             &
$\sigma(z_p)$             &
$\sigma(\cals)$             &
$\sigma(\calr)$            \\
\hline
0.00001 &  0.0052&0.0011&0.0266& 0.035& 0.034& 0.038 &0.035 &0.148&    0.050& 0.020& 0.020\\
0.00010 &  0.0037&0.0011&0.0216& 0.026& 0.032& 0.035 &0.035 &0.147&    0.050& 0.020& 0.020\\
0.00100 &  0.0033&0.0010&0.0182& 0.025& 0.025& 0.025 &0.034 &0.145&    0.048& 0.020& 0.020\\
0.01000 &  0.0030&0.0010&0.0167& 0.023& 0.021& 0.017 &0.033 &0.143&    0.041& 0.020& 0.020\\
0.10000 &  0.0028&0.0009&0.0160& 0.018& 0.019& 0.015 &0.031 &0.138&    0.026& 0.020& 0.020\\
1.00000 &  0.0022&0.0008&0.0132& 0.015& 0.016& 0.012 &0.026 &0.119&    0.018& 0.019& 0.020\\
\hline
0.00001 &  0.0051&0.0011&0.0264& 0.034& 0.034& 0.038 &0.035 &0.148&    0.049& 0.020& 0.020\\
0.00010 &  0.0037&0.0011&0.0215& 0.025& 0.031& 0.035 &0.034 &0.147&    0.047& 0.020& 0.020\\
0.00100 &  0.0032&0.0010&0.0180& 0.020& 0.024& 0.023 &0.032 &0.143&    0.037& 0.020& 0.019\\
0.01000 &  0.0029&0.0009&0.0159& 0.017& 0.019& 0.015 &0.030 &0.138&    0.027& 0.020& 0.015\\
0.10000 &  0.0026&0.0008&0.0122& 0.014& 0.016& 0.012 &0.026 &0.122&    0.019& 0.018& 0.008\\
1.00000 &  0.0019&0.0006&0.0067& 0.009& 0.010& 0.007 &0.020 &0.091&    0.011& 0.014& 0.003\\
\end{tabular}
\end{ruledtabular}
\end{table*}
\begin{table*}
\caption{\label{tab:varn}Parameter estimation errors for weak lensing
  combined with CMB: $\bar{n}$ is varied from $\bar{n}=5.0 \times 10^6
  \sr^{-1}$ to $\bar{n}=1.0 \times 10^9 \sr^{-1}$; $f_{sky}=0.01$. With
  a prior of 0.02 on $\calr$ and $\cals$ and 0.05 on $z_p$. Tomography
  is added in the second part of the table.}
\begin{ruledtabular}
\begin{tabular}{cccccccccccc}
$\bar{n}$&
$\sigma(\Omega_m h^2)$    &
$\sigma(\Omega_b h^2)$    & 
$\sigma(\Omega_\Lambda)$  &
$\sigma(\sigma_8)$        & 
$\sigma(n_s)$	          &
$\sigma(\alpha_s)$        &
$\sigma(\tau)$            &
$\sigma(T/S)$             &
$\sigma(z_p)$             &
$\sigma(\cals)$             &
$\sigma(\calr)$            \\ 
\hline
$5.0\times 10^6$ &0.0045&0.0011&0.0239& 0.032 &0.034 &0.038& 0.035& 0.149 &   0.050& 0.020& 0.020\\
$1.0\times 10^7$ &0.0035&0.0011&0.0194& 0.028 &0.030 &0.033& 0.035& 0.147 &   0.049& 0.020& 0.020\\
$5.0\times 10^7$ &0.0032&0.0010&0.0173& 0.024 &0.023 &0.022& 0.034& 0.145 &   0.046& 0.020& 0.020\\
$1.0\times 10^8$ &0.0032&0.0010&0.0170& 0.024 &0.022 &0.020& 0.034& 0.144 &   0.044& 0.020& 0.020\\
$5.0\times 10^8$ &0.0031&0.0010&0.0167& 0.023 &0.021 &0.017& 0.033& 0.143 &   0.041& 0.020& 0.020\\
$1.0\times 10^9$ &0.0030&0.0010&0.0166& 0.022 &0.021 &0.017& 0.033& 0.143 &   0.040& 0.020& 0.020\\
\hline
$5.0\times 10^6$ &0.0048&0.0011&0.0247& 0.033 &0.034 &0.038& 0.035& 0.149 &   0.050& 0.020& 0.020\\
$1.0\times 10^7$ &0.0036&0.0011&0.0197& 0.028 &0.030 &0.034& 0.035& 0.147 &   0.049& 0.020& 0.020\\
$5.0\times 10^7$ &0.0032&0.0010&0.0173& 0.021 &0.022 &0.021& 0.033& 0.143 &   0.040& 0.020& 0.019\\
$1.0\times 10^8$ &0.0031&0.0010&0.0168& 0.019 &0.021 &0.019& 0.032& 0.142 &   0.035& 0.020& 0.018\\
$5.0\times 10^8$ &0.0029&0.0009&0.0159& 0.017 &0.019 &0.016& 0.031& 0.138 &   0.027& 0.020& 0.015\\
$1.0\times 10^9$ &0.0028&0.0009&0.0148& 0.015 &0.018 &0.014& 0.030& 0.133 &   0.023& 0.019& 0.009\\
\end{tabular}
\end{ruledtabular}
\end{table*}
\begin{table*}
\caption{\label{tab:corrtab}Correlation matrix for parameter
  estimation errors for weak lensing combined with CMB. $\bar{n}=6.6e8$
  and $f_{sky}=0.01$. With a prior of
  0.02 on $\calr$ and $\cals$, and 0.05 on $z_p$.  The above-diagonal elements
  correspond to a survey without tomography; the below-diagonal elements
  correspond to a survey with tomography.}
\begin{ruledtabular}
\begin{tabular}{c|cccccccccccc}
                          &
$\Omega_m h^2$    &
$\Omega_\Lambda$  &
$\sigma_8$        & 
$n_s$	          &
$\alpha_s$        &
$z_p$             &
$\cals$           &
$\calr$           &
$\Omega_b h^2$    & 
$\tau$            &
$T/S$             \\
\hline
$\Omega_m h^2$    &
    \onediag &           -0.325 & -0.020 &  0.135  &-0.266  &-0.430  &-0.137 &  0.017 &  0.378 & -0.533  & 0.313\\  
$\Omega_\Lambda$  &
        -0.375   &  \onediag &     0.754 &  0.667  &-0.583  & 0.464  & 0.122 & -0.018 &  0.558 &  0.728  & 0.393\\
$\sigma_8$        & 
        -0.230  & 0.877     &      \onediag &       0.632  &-0.679  &-0.160  &-0.091 &  0.008 &  0.541 &  0.757  & 0.375\\
$n_s$	          &
         0.072  & 0.651 &  0.620  &  \onediag &     -0.774  & 0.104  & 0.022 & -0.002 &  0.788 &  0.393  & 0.794\\
$\alpha_s$        &
        -0.228  &-0.550 & -0.626 & -0.770   &  \onediag &    -0.076  & 0.026 &  0.005 & -0.716 & -0.433  &-0.752\\
$z_p$             &
        -0.386  & 0.818 &  0.570 &  0.444 & -0.455  & \onediag &     -0.052 &  0.020 &  0.092 &  0.126  & 0.096\\
$\cals$           &
        -0.156  & 0.078 & -0.217 & -0.028 &  0.106 & -0.045    & \onediag &         0.002 &  0.014 &  0.004  & 0.013\\
$\calr$           &
         0.086  & 0.158 &  0.386 &  0.221 & -0.204 & -0.243  & 0.083    &   \onediag &         -0.003 & -0.005  &-0.004\\
$\Omega_b h^2$    & 
         0.349  & 0.527 &  0.516 &  0.762 & -0.698 &  0.371  &-0.033  & 0.185     & \onediag &            0.372  & 0.673\\
$\tau$            &
        -0.680  & 0.726 &  0.732 &  0.329 & -0.321 &  0.572  &-0.049  & 0.176  & 0.311       &\onediag &             0.136\\
$T/S$             &
         0.294  & 0.355 &  0.357 &  0.786 & -0.756 &  0.266  &-0.028  & 0.133  & 0.650 &  0.073    & \onediag \\
\end{tabular}
\end{ruledtabular}
\end{table*}

\begin{table*}
\caption{\label{tab:varfsky2}Parameter estimation errors for weak lensing
  combined with CMB: Here we have fixed $\calr$, $\cals$ and
  $z_p$ by imposing priors of width $10^{-4}$ on these parameters; $f_{sky}$ is varied from
  $10^{-5}$ to 1.0; $\bar{n}=6.6 \times 10^8 \sr^{-1}$.  
  Tomography is added in second part of the table.} 
\begin{ruledtabular}
\begin{tabular}{cccccccccccc}
$f_{sky}$&
$\sigma(\Omega_m h^2)$    &
$\sigma(\Omega_b h^2)$    & 
$\sigma(\Omega_\Lambda)$  &
$\sigma(\sigma_8)$        & 
$\sigma(n_s)$	          &
$\sigma(\alpha_s)$        &
$\sigma(\tau)$            &
$\sigma(T/S)$             &
$\sigma(z_p)$             &
$\sigma(\cals)$             &
$\sigma(\calr)$            \\
\hline
0.00001 &  0.0050&0.0011&0.0260 &0.034 &0.034& 0.038 &0.035& 0.148 &  0.0001&0.0001&0.0001\\
0.00010 &  0.0034&0.0011&0.0206 &0.025 &0.032& 0.035 &0.035& 0.147 &  0.0001&0.0001&0.0001\\
0.00100 &  0.0029&0.0010&0.0169 &0.023 &0.025& 0.025 &0.034& 0.145 &  0.0001&0.0001&0.0001\\
0.01000 &  0.0027&0.0010&0.0145 &0.022 &0.021& 0.017 &0.033& 0.142 &  0.0001&0.0001&0.0001\\
0.10000 &  0.0026&0.0009&0.0105 &0.016 &0.017& 0.013 &0.027& 0.132 &  0.0001&0.0001&0.0001\\
1.00000 &  0.0022&0.0007&0.0044 &0.007 &0.011& 0.008 &0.019& 0.104 &  0.0001&0.0001&0.0001\\
\hline
0.00001 &  0.0049&0.0011&0.0258 &0.033 &0.034& 0.038 &0.035& 0.148 &  0.0001&0.0001&0.0001\\
0.00010 &  0.0033&0.0011&0.0204 &0.024 &0.031& 0.035 &0.034& 0.147 &  0.0001&0.0001&0.0001\\
0.00100 &  0.0029&0.0010&0.0146 &0.019 &0.023& 0.022 &0.031& 0.140 &  0.0001&0.0001&0.0001\\
0.01000 &  0.0027&0.0008&0.0069 &0.010 &0.016& 0.013 &0.023& 0.130 &  0.0001&0.0001&0.0001\\
0.10000 &  0.0024&0.0007&0.0024 &0.003 &0.012& 0.008 &0.019& 0.112 &  0.0001&0.0001&0.0001\\
1.00000 &  0.0019&0.0006&0.0008 &0.001 &0.007& 0.004 &0.017& 0.085 &  0.0001&0.0001&0.0001\\
\end{tabular}
\end{ruledtabular}
\end{table*}

\section{Discussion}

In accord with previous studies, we find that when weak lensing surveys are combined with the CMB results, the
uncertainties on the cosmological parameters are reduced.  The improvement can be up to an order of magnitude,
depending on the size and depth of the survey (see Table \ref{tab:varfsky} and Table \ref{tab:varn}).  It is well
known that weak lensing and CMB have different types of degeneracies in their parameters which are nicely broken when
combined together.  In particular, weak lensing does not suffer from the well-known angular diameter distance
degeneracy; and it probes smaller comoving scales than the CMB, which means that it has a different degeneracy 
direction in the $(n_s,\alpha_s)$ plane.  The improvements from including weak lensing are 
especially notable for $\sigma_8$, $\Omega_m h^2$ and $\Omega_{\Lambda}$.

Motivated by recent discussion concerning the running of the primordial power 
spectrum $\alpha_s$, we include it as 
 a parameter in the weak lensing analysis. We find that for small surveys, modest improvement is obtained for $n_s$
and $\alpha_s$. Our Reference Survey can reduce $\sigma(n_s)$ and $\sigma(\alpha_s)$ by roughly a factor of two.  
For surveys within the near future (see Refs. in
\S\ref{sec:intro}), weak lensing can be used to provide complementary 
constraints to detect a
possible running spectral index, but may not be able to verify the 
results obtained by combined CMB+Lyman-$\alpha$ forest analysis, which should 
give another factor of 2-4 lower $\sigma(\alpha_s)$. A detailed comparative study 
of the constraints from the two probes is left for future work.
We find that tomography improves in particular the uncertainty on $\sigma_8$
and $\Omega_\Lambda$. For
the Reference Experiment, we do not find large improvements in the other parameters from tomography except for the
most ambitious surveys. This can be attributed to the strong parameter degeneracies present even when tomography is
used (see Table \ref{tab:corrtab}), most notably the degeneracy between $\Omega_\Lambda$ and $z_p$.  If the source
redshift distribution is known accurately from a spectroscopic redshift survey, then this degeneracy is lifted, and
tomography becomes a powerful tool for measuring cosmological parameters, including $\Omega_\Lambda$, $n_s$ and
$\alpha_s$ (see Table \ref{tab:varfsky2}).  A comparison of Tables \ref{tab:varfsky} and \ref{tab:varfsky2} shows
that the parameter constraints from tomography are more sensitive to the systematics parameters than the constraints
without tomography, hence the benefits of tomography can only be fully realized if systematic errors
are tightly controlled.

We expanded the usual weak lensing parameter space to include two calibration parameters in addition to the
characteristic redshift of source galaxies.
We hope that the present analysis will encourage weak lensing observers to expand their likelihood analyses to
include a parameterization of systematic errors.  It is already common practice to report results marginalized over
the characteristic source redshift \cite{2003PhRvL..90v1303C,2002ApJ...577..595H,2002ApJ...572...55H} or to treat it
as a systematic to be added in quadrature to statistical errors \cite{2003AJ....125.1014J}. Ultimately it would be
desirable to include not just the calibration factors and characteristic source redshift but also the full redshift
distribution of sources, intrinsic alignments, etc.  Additional data (such as spectroscopic redshifts) and detailed
analysis and simulations may be required in order to constrain some of these parameters.  However, in some cases it
is possible to obtain information about the systematics parameters from the data itself.  For example, the nonlinear
portion of the convergence power spectrum provides joint constraints on the shear calibration biases $\cals$ and
$\calr$ and the cosmological parameters.

We have shown, at least at the level of our Reference Survey, that 
{\sc halofit} provides a sufficiently good fit to
$N$-body simulations for use in parameter forecasting studies.  The
largest discrepancy is $25$\% in the error on $n_s$ when tomography 
is included, but most of the error estimates agree to better than 10\%.
Note that this does not imply that the nonlinear mapping is sufficiently
accurate for analysis of Reference Survey data, since the best-fit values
of cosmological parameters can be significantly affected by small errors
in theoretical predictions even if the fractional change in the Fisher
matrix is small.

In summary, we have shown that weak lensing, supplemented with the one-year \WMAP data, has the potential to very
precisely measure the running of the scalar spectral index.  This would provide a third measurement of $\alpha_s$ in
addition to those using the Lyman-$\alpha$ forest and/or galaxy power spectrum, and the precision measurement of the
high-$\ell$ CMB power spectrum expected from {\slshape Planck}. Weak lensing observations are currently progressing
rapidly and the data required to significantly improve \WMAP constraints on $\alpha_s$ should be available in the
foreseeable future. By providing an additional and mostly independent 
measurement of the value of $\alpha_s$ (and
more generally the scalar power spectrum) obtained by these other methods, weak lensing will help us to understand
the spectrum of primordial scalar fluctuations in the universe and thus provide valuable information on the mechanism
of their generation.

\acknowledgments
M.I. thanks K.~F. Huffenberger and R.~E. Smith for useful comments on
the non-linear mapping procedures. We
thank Alexey Makarov for providing the CMB covariance
matrix of current data. We thank Paul
Bode for the TPM code. The simulations were performed at the National
Center for Supercomputing Applications (NCSA).

M.I. acknowledges the support of 
the Natural Sciences and Engineering Research Council of Canada
(NSERC) PDF Fellowship Program. C.H. acknowledges the support of the National Aeronautics
and Space Administration (NASA) Graduate Student Researchers Program (GSRP).
U.S. acknowledges support from Packard
and Sloan foundations, NASA NAG5-11983 and NSF CAREER-0132953.

\end{document}